# Development of Record and Management Software for GPS/Loran Measurements

Woohyun Kim[1], Pyo-Woong Son[2], Joon Hyo Rhee[3], and Jiwon Seo[1*]

[1] School of Integrated Technology, Yonsei University,
Incheon, 21983, Korea (crimy00, jiwon.seo@yonsei.ac.kr)
[2] Korea Research Institute of Ships and Ocean Engineering,
Daejeon, 34103, Korea (pwson@kriso.re.kr)
[3] Korea Research Institute of Standards and Science,
Daejeon, 34113, Korea (jh.rhee@kriss.re.kr)

* Corresponding author

**Abstract**: In this paper, a software implementation that records Global Positioning System (GPS) and long-range navigation (Loran) measurement data output from an integrated GPS/Loran receiver and organizes them based on time is proposed. The purpose of the developed software is to collect measurements from multiple Loran transmitter chains for performance analysis of navigation methods using Loran, and to organize the data based on time to make it easy to use them. In addition, GPS measurements are also collected and managed as ground truth data for performance analysis. The implemented software consists of three modules: recording, classification, and conversion. The recording module records raw text data streamed from the receiver, and the classification module classifies the recorded text data according to the message format. The conversion module parses the classified text data, sorts GPS and Loran measurements based on timestamp, and outputs them according to the software platform of the user to analyze the measurements. Each module of the software runs automatically without user intervention. The functionality of the implemented software was verified using GPS and Loran measurements collected over 24 h from an actual integrated GPS/Loran receiver.

## 1. INTRODUCTION

The global navigation satellite systems (GNSSs) [1-4], which utilize navigation signals that are transmitted from satellites, are widely employed in various fields owing to the advantages of highly accurate and easily accessible positioning, navigation, and timing (PNT) capabilities [5-8]. Systems operating across the globe include Global Positioning System (GPS) of the United States, GLONASS of Russia, Galileo in Europe, and BeiDou in China. However, since the altitude of the navigation satellites is very high, the strength of the navigation signal received from the ground is very weak. Owing to this, GNSS has drawbacks such as difficulty in receiving signals in an urban environment [9-11] and vulnerability to radio frequency interference (RFI) [12, 13] and ionospheric anomalies [14-17].

As a method for mitigating the effects of RFI, GNSS controlled reception pattern antennas (CRPAs) have been studied [18-21]. However, GNSS CRPAs are larger and costlier than conventional GNSS receivers using a single-element antenna, and their effectiveness is limited in large-scale high-power jamming attacks [22].

Meanwhile, long-range navigation (Loran) [22-24], which utilizes terrestrial radio signals for navigation, is significantly more robust to RFI than GNSS. Since Loran uses a high-power navigation signal on the ground, the strength of signal received by users is considerably higher than that of GNSS, making it resistant to RFI. Because of this characteristic, Loran can be used as an alternative navigation system in environments where GNSS cannot be used. However, the accuracy of the navigation solution of Loran is lower than that of GNSS, making it difficult to replace GNSS completely. Therefore, studies have been conducted to improve the accuracy of Loran [22-27], and enhanced Loran (eLoran) was also proposed [28].

Conventional Loran system utilizes a chain of two or more secondary transmitters that are time-synchronized with one master transmitter. The group repetition interval (GRI), which is the time interval between the reoccurrence of the master transmission, is designated differently for each chain so that it can be distinguished from each other. A typical Loran receiver can retrieve its position only with measurements obtained from transmitters in the same chain, thereby limiting the positioning accuracy and availability. However, it is feasible to improve the positioning performance using measurements obtained from multiple chains by utilizing the multichain Loran positioning algorithm [22].

As different methods to improve the performance of Loran were developed, the need for a platform to evaluate the performance of these methods has also increased. Such a platform should be able to collect measurements from multiple Loran transmitter chains and organize data based on timestamp to make it convenient to analyze. GNSS measurements, which are the ground truth data for performance evaluation, should also be collected.

Therefore, we propose a software implementation that can collect measurement outputs of GPS and Loran from the integrated GPS/Loran receiver, manage data based on time, and execute without user intervention through an automated execution process. Section 2 of this paper describes the system configuration of the proposed software, and Section 3 describes the implementation of the proposed software. Finally, Section 4 provides the paper conclusion.

## 2. SYSTEM CONFIGURATION

The aim of the proposed software is to collect raw data output from an integrated GPS/Loran receiver, convert it into a data format required for analysis, and organize it according to time. To this end, the functions of the proposed software are separated into three modules: recording, classification, and conversion. Fig. 1 depicts this graphically.

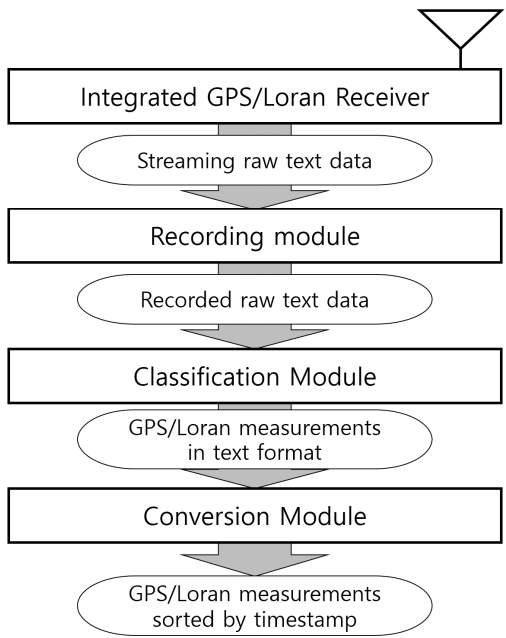

Fig. 1 Overall process of the proposed software implementation.

The first module of the software, recording module, records raw data output from the integrated GPS/Loran receiver. Since the raw data output from a typical receiver consists of streaming text, the module focuses on recording the data without loss. Therefore, there is no special processing of the recorded data, and the measurements of GPS and Loran are not distinguished.

The classification module extracts messages from raw text data and classifies them by type. The integrated GPS/Loran receiver outputs data by constructing a message in a certain format, such as NMEA 0183 [29] or proprietary format. Since the recording module records output data from the receiver regardless of the message format, it cannot check the types of messages stored. Therefore, the classification module reads the raw text data, checks which messages are recorded, and classifies them into separate text data. The classification module uses the heading information of each line of raw text data for the classification of GPS and Loran measurements.

The final module, the conversion module, accepts the text data organized by the classification module, parses it according to the message format, and converts it for use in analysis. The converted GPS and Loran measurements are sorted by timestamp so that the user can analyze them. The conversion module converts the measurement data according to the programming language or platform used by the user for data analysis, such as Python or MATLAB, and can be directly used for analysis without any additional process.

Each module in the software runs independently, reducing problems caused by inter-module communication. Furthermore, the execution of each module is automated, enabling reliable data collection without user intervention.

## 3. IMPLEMENTATION

The proposed software solution was implemented using a Windows PC equipped with a reelektronika LORADD eLoran receiver. The receiver can simultaneously receive GPS and Loran signals using an integrated antenna, and the processing results of the receiver are output to the PC using the NMEA standard and proprietary format.

In this study, we assumed that the user analyzes GPS and Loran measurement data using MATLAB and implemented the proposed software solution accordingly. The recording and classification modules were implemented using Python because they only need the ability to store and process text, and do not depend on the platform used for the analysis of GPS and Loran data. On the other hand, the conversion module was implemented using MATLAB so that its output can be a MATLAB structure.

Each module of the proposed software was implemented to automatically initiate its process every 24 h. The recording module records the data output from the receiver for 24 h, and the classification and conversion modules are executed once every 24 h.

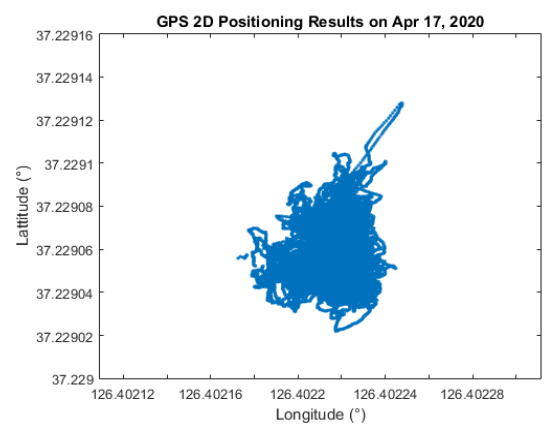

Fig. 2 Positioning results from GPS measurements.

To validate the functioning of the implemented software, the data was collected for 24 h on April 17, 2020 using an antenna located on the rooftop of a building at Yonsei University, Incheon, Korea. Fig. 2 shows positioning results obtained from GPS measurements in 24 h. Fig. 3 shows the signal-to-noise ratio (SNR) of signals received from a Loran transmitter (9930M) for 24 h. From these two results, it is confirmed that the developed software can collect and manage GPS and Loran measurements successfully. This software was used in the multichain Loran accuracy study in [30].

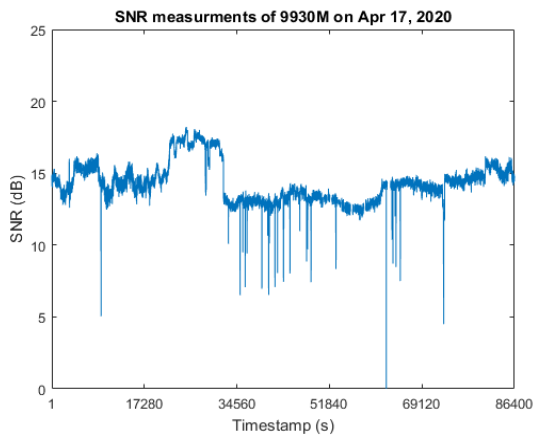

Fig. 3 Signal-to-noise ratio (SNR) of signals received from Loran transmitter 9930M.

## 4. CONCLUSION

In this study, we implemented software that records GPS and Loran measurement data output from an integrated GPS/Loran receiver and organizes them based on time. The implemented software consists of three modules: recording, classification, and conversion. The recording module records raw text data streamed from the receiver, and the classification module classifies the recorded text data according to the message format. The conversion module parses the classified text data, sorts GPS and Loran measurements based on timestamp, and outputs them according to the platform used by the user to analyze the measurements. Each module of the software runs automatically without user intervention. The functionality of the implemented software was demonstrated using GPS and Loran measurements collected over 24 h from an actual integrated GPS/Loran receiver.

## ACKNOWLEDGEMENT

This research was a part of the project titled "Development of enhanced Loran system," funded by the Ministry of Oceans and Fisheries, Korea.